\documentclass[prb,aps,superscriptaddress,reprint]{revtex4-1}
\usepackage{graphicx}
\usepackage{amsmath}
\usepackage{float}
\usepackage{amssymb}
\usepackage[makeroom]{cancel}
\usepackage[english]{babel}
\usepackage{color}

\begin{document}
\title{Renormalization of the Majorana bound state\\ decay length in a perpendicular magnetic field}

\author{M. P. Nowak}
\affiliation{AGH University of Science and Technology, Academic Centre for Materials and Nanotechnology, al. A. Mickiewicza 30, 30-059 Krakow, Poland}

\author{P. W{\'o}jcik}
\affiliation{AGH University of Science and Technology, Faculty of Physics and Applied Computer Science, al. A. Mickiewicza 30, 30-059 Krakow, Poland}
\date{\today}

\begin{abstract}
Orbital effects of a magnetic field in a proximitized semiconductor nanowire are studied in the context of the spatial extent of Majorana bound
states. We develop analytical model that explains the impact of concurring effects of paramagnetic coupling of the nanowire bands via the kinetic
energy operator and spin-orbit interaction on the Majorana modes. We find, that the perpendicular field, so far considered as to be detrimental to the
Majorana fermion formation, is in fact helpful in establishing the topological zero-energy-modes in a finite system due to significant decrease in the
Majorana decay length.
\end{abstract}

\maketitle

\section{Introduction}
Topological superconductivity created by a combination of the spin-orbit (SO) coupling and the Zeeman effect in a low-dimensional
semiconductor-superconductor heterostructures \cite{oreg_helical_2010, sau_generic_2010} is foreseen as prospective in the quantum computation that
exploits manipulation of Majorana bound states (MBSs) \cite{kitaev_fault-tolerant_2003}. The elementary braiding operation of Majoranas requires at
least a three terminal junction. In the pursuit of topological quantum gates creation \cite{alicea_non-abelian_2011, heck_coulomb-assisted_2012,
hyart_flux-controlled_2013} multiterminal networks of MBSs are realized experimentally in a form of nanowire crosses \cite{plissard_formation_2013,
fadaly_observation_2017} and hashtags \cite{gazibegovic_epitaxy_2017}. In those structures orientation of the magnetic field perpendicular to the
substrate is preferable as it allows to induce the topological phase in all the nanowire branches owing to perpendicular orientation of the field with
respect to the SO coupling direction.

In previous experimental research, large in-plane, type-II superconducting electrodes were sputtered after the nanowire deposition on the
substrate \cite{mourik_signatures_2012, deng_anomalous_2012, finck_anomalous_2013, churchill_superconductor-nanowire_2013, chen_experimental_2017},
prohibiting application of the perpendicular magnetic field due to the vortex formation in the superconductor slab. 
This obstacle has been recently overcome by the realization of 2DEG \cite{shabani_two-dimensional_2016} and nanowire \cite{krogstrup_epitaxy_2015,
gazibegovic_epitaxy_2017} hybrid devices in a bottom-up synthesis with a thin Aluminum shell. This guarantees a pristine interface between
superconductor and semiconductor materials \cite{kjaergaard_transparent_2017}, ensures the hard superconducting gap \cite{chang_hard_2015} and allows
for observation of a long-predicted conductance doubling in a proximitized quantum point contact \cite{kjaergaard_quantized_2016} or $2e^2/h$
quantization of Majorana modes \cite{zhang_quantized_2017}. Most importantly, the exploitation of the thin superconducting shell enables arbitrary
alignment of the magnetic field without destroying the superconductivity \cite{albrecht_exponential_2016} and by that opens a wide perspective for the
topological quantum gate creation.

Studies that took into account effects of the magnetic field beyond sole-Zeeman-interaction focused on the impact of the field orientation on the
phase boundaries \cite{lim_magnetic-field_2012, osca_majorana_2015, nijholt_orbital_2016, kiczek_influence_2017} and painted somewhat discouraging
picture of topological phase instability due to the gap closing in tilted magnetic field \cite{lim_magnetic-field_2012}. It has been shown however
that this is rather an effect of numerical treatment of the vector potential \cite{osca_majorana_2015}, and in fact the topological phase can still
exist in perpendicular orientation of the field. In this work, exploiting analytical treatment of the orbital effects of the perpendicular magnetic
field we address problem of the real-space extent of Majorana modes, crucial in the light of topological protection of MBSs in a finite-system quantum
gates.

Very recently O. Dmytruk and J. Klinovaja \cite{dmytruk_suppression_2017} pointed out that the diamagnetic effect of the magnetic field oriented {\it
along} the radially symmetric nanowire acts as an effective chemical potential that reduces the energy oscillations of the overlapping MBSs. This is
true only for the parallel magnetic field orientation when the paramagnetic effect does not couple the orbital and spin degrees of freedom. As we
show, in general case when the magnetic field has a component perpendicular to the nanowire axis or, as in our case, it is simply perpendicular to the
nanowire, the paramagnetic coupling of modes with different orbital excitation and opposite spin renormalizes not only the chemical potential but also
the effective mass and the SO coupling constant. As a result both the diamagnetic and paramagnetic terms in the kinetic energy operator contribute to
the decrease of the spatial extent of the Majorana modes. On the other hand the magnetic field entering through momentum operator in the SO coupling
Hamiltonian has distinct and detrimental effects on the reduction of the decay length, which is crucial for current generation of devices where SO
interaction is particularly strong \cite{van_weperen_spin-orbit_2015, heedt_signatures_2017, kammhuber_conductance_2017}. We find surprising
equalization of the above mentioned effects that results in the decay length comparable, but still less than in the case without the orbital effects
of the magnetic field.

\section{Theoretical model}
We start by writing down Bogoliubov-de Gennes Hamiltonian for a proximitized semiconductor nanowire,
\begin{equation}
\begin{split}
H = \left( \hbar^2\mathbf{k}^2/2m^*-\mu\right)\sigma_0\tau_z+\Delta\sigma_0\tau_x\\
+\alpha(\sigma_xk_y-\sigma_yk_x)\tau_z+E_z\sigma_z\tau_0.
\end{split}
\label{2DHamiltonian}
\end{equation}
where $\alpha$ is the Rashba SO coupling constant, $E_z = \frac{1}{2}g\mu_BB$ is the Zeeman term, $B$ is the external magnetic field oriented in the
$z$-direction, perpendicular to the nanowire plane and $\Delta$ is the effective induced pairing potential. $\sigma_i$ and $\tau_i$ with $i=x,y,z$ are
the Pauli matrices acting on spin- and particle-hole degrees of freedom, respectively. The uniform $\Delta$ in Eq. (\ref{2DHamiltonian}) corresponds to a system in the weak-coupling regime \cite{peng_strong_2015}, where the superconductor-semiconductor interface is non-transparent or to the presence of low-energy modes with minimum in the interface transparency \cite{sticlet_robustness_2017}. The orbital effects of the
magnetic field are included through the canonical momentum, $\mathbf{k}=-i\nabla+e\mathbf{A}/\hbar\cdot\tau_z$ with the vector-potential in Lorentz
gauge $\mathbf{A}=[-yB,0,0]$.

\begin{figure}[h!]
\center
\includegraphics[width = 8.5 cm]{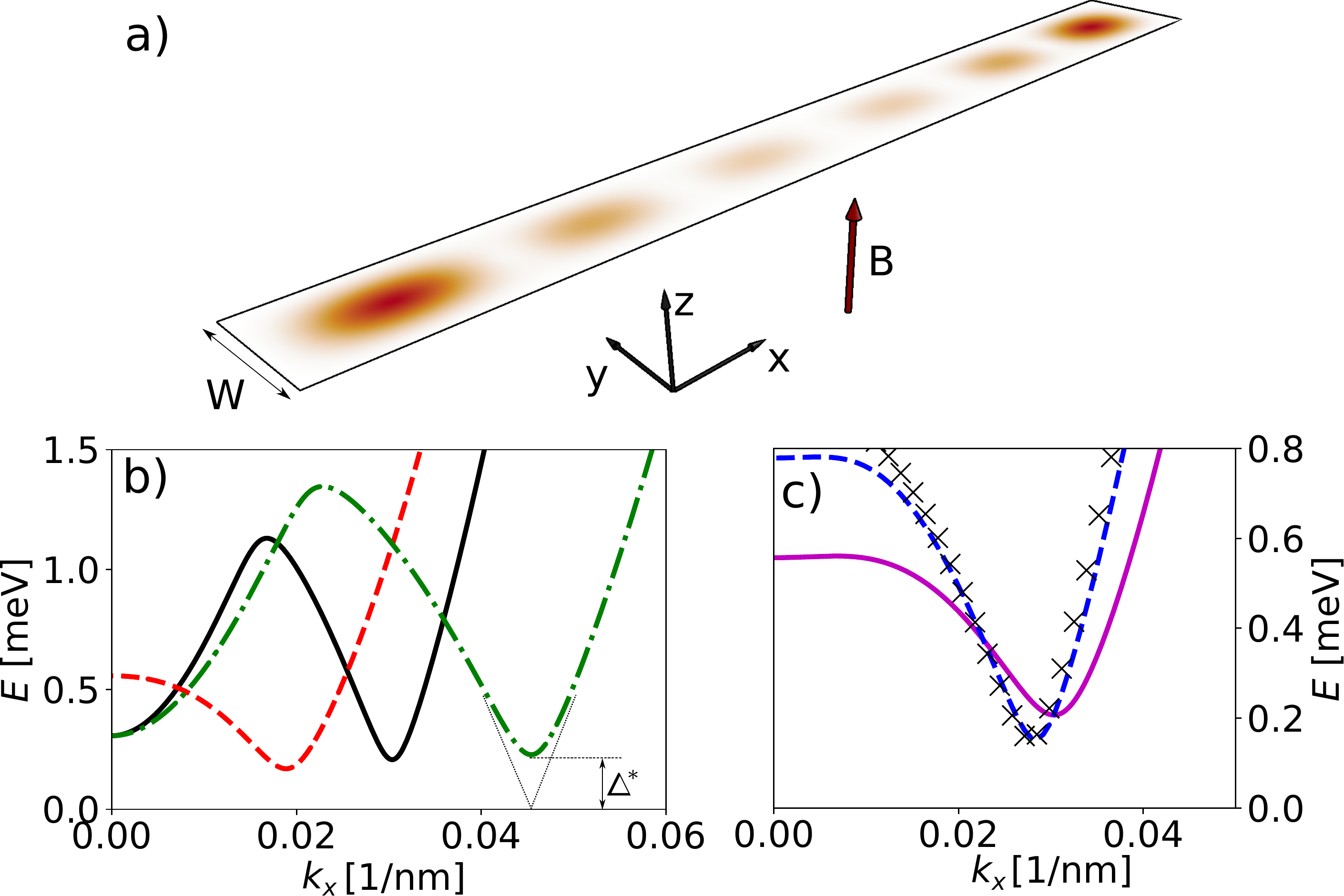}
\caption{
a) Two dimensional crossection of the proximitized semiconductor nanowire subjected to the magnetic field $B$ perpendicular to the nanowire axis. The
colormap shows an exemplary density distribution of the Majorana bound state. 
b) Dispersion relations $E(k)$ obtained in the model of Eq. (\ref{analytical_energy}) without the orbital effects (solid black), with the
kinetic-paramagnetic term (green dashed) and with the diamagnetic term (red dashed). Their joint effect is presented in panel c) with the violet curve.
In c) the blue dashed curve corresponds to the results with the canonical momentum included also in the SO coupling Hamiltonian (SO-paramagnetic
term). The black crosses are the results obtained in numerical calculations. 
Results obtained for $\mu = 3$ meV, $\alpha = 50$ meVnm, and $B=0.7$ T.}
\label{dispersions}
\end{figure}

It has been demonstrated that the most convenient situation for realization of MBSs is when a single band of transverse quantization is occupied
\cite{nijholt_orbital_2016}. We therefore focus on the low electron density regime where the system is tuned such the phase transition occurs due to a
single band crossing the Fermi level. Since we are specifically interested in the orbital effects of the magnetic field, for simplicity and without
loss of generality, we decouple the spatial degree of freedom collinear with the magnetic field $z$-direction and consider the Hamiltonian
(\ref{2DHamiltonian}) in the $x-y$ plane. Figure \ref{dispersions}~a)
shows the schematic illustration of the considered nanowire oriented along $x$ direction with the width $W$ along $y$ (transverse) direction.

In contrary to the diamagnetic term ($\sim y^2 B^2$) which solely shifts energies of the electronic states, the paramagnetic effect ($\sim k_x yB$)
originating from both the kinetic energy (kinetic-paramagnetic term) and the SO interaction (SO-paramagnetic term), couples the orbital degrees of
freedom in the $y$ direction via the wave vector $k_x$. For Majorana fermions in the ground state of transverse quantization, the most significant
effect comes from the coupling with the first excited state. For this purpose we write the Hamiltonian (\ref{2DHamiltonian}), in the basis of two
lowest eigenstates of infinite quantum well of width $W$, centered at $y=0$, $\Psi_n(y) = \sqrt{2/W}\sin[n\pi(y+W/2)/W]$. We obtain 
\begin{equation}
H=\left(
\begin{matrix}
H_{11} & H_{12} \\
H_{21} & H_{22}
\end{matrix}\right),
\label{two_band_model}
\end{equation}
with
\begin{eqnarray}
\label{h_11}
&H&_{11(22)}=H_{1D} + (E_{1(2)} + E_{1(2)}^{\mathrm{dia}})\sigma_0\tau_z,\\
\label{h_12}
&H&_{12}=H_{21}=\varepsilon_{p}k_x\sigma_0\tau_0 + E^{SO}_p\sigma_y\tau_0 + E_{\bot}^{SO}\sigma_x\tau_z,
\end{eqnarray}
where $E_n= n^2\pi^2\hbar^2/2mW^2$ is the energy of orbital excitation in the $y$-direction,
$E_{n}^{\mathrm{dia}}=\langle \Psi_n |y^2| \Psi_n \rangle e^2B^2/2m$ is the diamagnetic term in the $n$'th subband [with 
$\langle \Psi_1 |y^2| \Psi_1 \rangle=(\pi^2-6)W^2/12\pi^2$ and $\langle \Psi_2 |y^2| \Psi_2 \rangle = (2\pi^2-3)W^2/24\pi^2$]
that acts as the extra chemical potential being different for each band. The parameter $\varepsilon_{p}=-\langle \Psi_1 | y| \Psi_2 \rangle e B \hbar
/m =16W e B \hbar / 9 m \pi^2$ results from the kinetic-paramagnetic effect that due to parity of the transverse modes is 
non-zero only in the off-diagonal submatrices ($H_{12}, H_{21}$). It mixes the transverse modes with magnitude proportional to the wave-vector $k_x$.
In Eq. (\ref{h_12}), $E^{SO}_p=\langle\Psi_1 |y|\Psi_2 \rangle\alpha e B/\hbar = -16W\alpha e B/ 9 \hbar \pi^2$ corresponds to the substitution of
the canonical momentum in the SO Hamiltonian that accounts for mixing bands with different transverse excitation
and spin along the $z$-direction. $E_{\bot}^{SO} = i\alpha \langle\Psi_1 |\partial/\partial_y |\Psi_2\rangle =  8i\alpha / 3W$ is a
correction due to the transverse part of the Rashba Hamiltonian. In Eq.~(\ref{h_11}) $H_{1D}$ is the one-dimensional Hamiltonian along the wire axis
\begin{equation}
\begin{split}
H_{1D}= \left( \frac{\hbar^2k_x^2}{2 m^*} - \mu\right) \sigma_0\tau_z+\Delta\sigma_0\tau_x \\
-\alpha k_x\sigma_y\tau_z+E_Z\sigma_z\tau_0.
\end{split}
\label{1DHamiltonian}
\end{equation}

In order to solve the problem analytically we use the fact that $E_2$ is the largest energy in the system. Then, the Hamiltonian $H_{22}$ can be
further simplified to diagonal form neglecting $\Delta$, $\alpha$, $E_z$ dependencies [for details see the Appendix A]. As we will see further, this
approximation works well and the calculated dispersions $E(k)$ remain in a good agreement with results obtained from the full numerical
diagonalization of the Hamiltonian (\ref{two_band_model}). Using the folding-down transformation,
\begin{equation}
\mathcal{H}(E)=H_{11}-H_{12}(H_{22}-E)^{-1}H_{21},
\end{equation}
the $8\times 8$ Hamiltonian (\ref{two_band_model}) can be reduced into the $4\times 4$ effective Hamiltonian. As we derived, it has the form of
$H_{1D}$ Eq. (\ref{1DHamiltonian}) with the renormalized effective mass $\tilde{m}^*$, chemical potential $\tilde{\mu}$ and SO coupling constant
$\tilde{\alpha}$ given by the formulas:
\begin{equation}
\label{factorm}
 \frac{1}{\tilde{m}^*}=\frac{1}{m^*}-\frac{2\varepsilon_{p}^2}{\hbar ^2 E_2},
\end{equation}
\begin{equation}
\label{factormu}
 \tilde{\mu}=\mu-E_1-E_{1}^{dia}+\frac{(E^{SO}_p-E_{\bot}^{SO})^2}{E _2},
\end{equation}
\begin{equation}
\label{factoralpha}
 \tilde{\alpha}= \alpha  +2\frac{E^{SO}_p\varepsilon_{p}}{E _2}.
\end{equation}
By diagonalization of the renormalized Hamiltonian (\ref{1DHamiltonian}) we find the energies
\begin{small}
\begin{equation}
\label{analytical_energy}
\begin{split}
 &E(k_x)=\\ &\pm \sqrt{\tilde{\xi} _1^2+E_Z^2+\tilde{\Delta} _{SO}^2+\Delta ^2 
 \pm 2 \sqrt{E_Z^2\tilde{\xi}_1^2+\tilde{\Delta} _{SO}^2\tilde{\xi}_1^2+\Delta^2E_Z^2} },
\end{split}
\end{equation}
\end{small}
where $\tilde{\xi}_1=\hbar^2 k_x^2/2\tilde{m}^*-\tilde{\mu}$ and $\tilde{\Delta}_{SO}=\tilde{\alpha}k_x$.

\section{Results}
 \subsection{Low-energy analysis}
The real space extent of MBSs with the wave-function $\sim e^{-x/\xi}$ is determined by the decay length $\xi$. For wires with length $L
\bcancel{\gg} \xi$ the modes overlap and their energies are shifted away from zero impairing the topological protection \cite{sarma_majorana_2015}.
 The localization length $\xi$ can be determined from the dispersion $E(k)$ of the translation invariant system by considering the
evanescent modes at zero energy. Away from the topological transition, $\xi$ is characterized by properties of the gapped Dirac cones at $k\neq 0$
with $\xi = \hbar \nu / \Delta^*$ in analogy to standard superconducting coherence length, where $\nu=dE/dk$ is the Fermi velocity and $\Delta^*$ is
the induced gap [see Fig. \ref{dispersions} b)]. 

Let us consider the band-structure given by Eq. (\ref{analytical_energy}) taking the parameters for the state-of-the-art structures made
on InSb nanowires covered by the Al shell \cite{gazibegovic_epitaxy_2017, fadaly_observation_2017, zhang_quantized_2017}. We adopt the effective
electron mass $m^* = 0.014$, induced gap $\Delta = 250\; \mu eV$ and considerable g-factor of $g = -51$ \cite{winkler_orbital_2017}. The SO
interaction strength has been recently probed in various types of measurements and proved to be quite diverse, ranging from $\alpha \simeq 20$ meVnm
in the spin-qubit manipulation experiments \cite{nadj-perge_spectroscopy_2012},
through $\alpha=50 - 100$ meVnm for the weak antilocalization measurements \cite{van_weperen_spin-orbit_2015}, up to extremely high values 
$\alpha=120$ meVnm or $\alpha=266$ meVnm in the transport experiments probing the helical gap \cite{heedt_signatures_2017,
kammhuber_conductance_2017}. Therefore, in the following we analyze results for a range of $\alpha$ values. We take the width of the wire $W=104$ nm [see Appendix B for the analysis of the width dependence]. 

In Fig. \ref{dispersions} b) we plot the positive-energy bands with the orbital effects of the
magnetic field neglected (black solid curve) and with the kinetic-paramagnetic term (green dashed curve) and diamagnetic term (red dashed curve)
included. By the analysis of these curves together with the formulas (\ref{factorm}), (\ref{factormu}) we conclude that the diamagnetic effect lowers
the chemical potential, while the paramagnetic effect widens the dispersion relation $E(k_x)$ due to rescaling of the effective mass. In turn, both
these effects decrease the slope of the cones at $k \neq 0$ decreasing the decay length $\xi$. Their joint outcome is depicted in Fig.
\ref{dispersions} c) with the violet curve. Inspecting Eqs. (\ref{factormu}, \ref{factoralpha}) we note that the paramagnetic term through the SO
interaction Hamiltonian {\it counteracts} the rescaling of the chemical potential and decreases the induced gap due the reduction of the SO coupling
constant [see the blue dashed curve in Fig. \ref{dispersions} c)].

\begin{figure}[ht!]
\center
\includegraphics[width = 7.5cm]{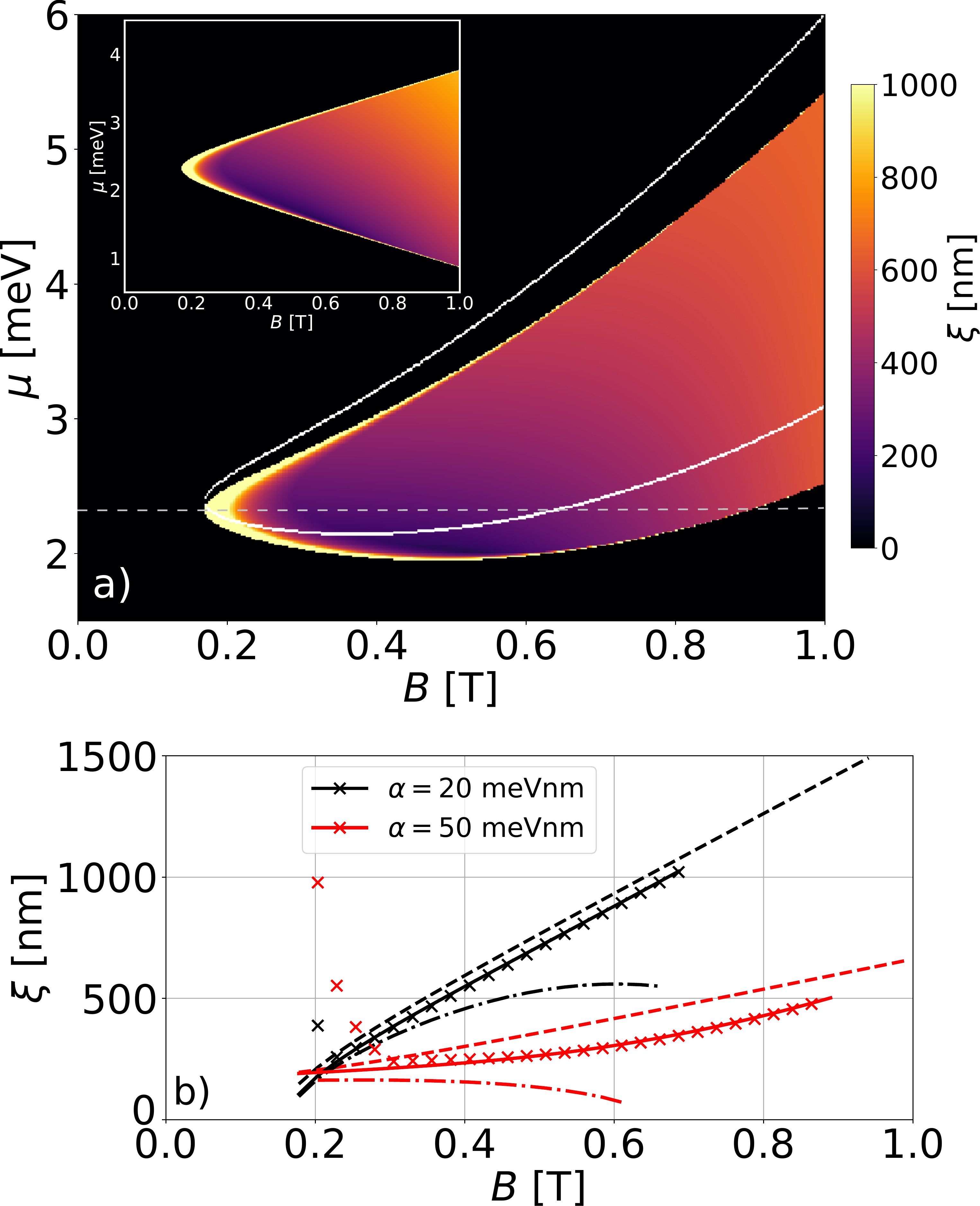}
\caption{a) Majorana bound state decay length obtained in the renormalized model of Eq. (\ref{1DHamiltonian}) for $\alpha = 50$ meV nm. The
white contour presents the topological phase boundary $E_z^2 > \Delta^2 + \tilde{\mu}^2$ when the orbital effects of the magnetic field are included
solely through the kinetic energy operator. The inset shows the decay length with the orbital effects neglected. b) Comparison of the decay
length calculated from Eq. (\ref{xi_theory}) without the orbital effects (dashed curves), with the kinetic-paramagnetic and the diamagnetic contributions included (dot-dashed
curves) and with the orbital effects included in both the kinetic energy operator and the SO coupling Hamiltonian (solid curves). The crosses present the decay length obtained from the model of Eq. (\ref{1DHamiltonian}). Results for $\alpha
= 20$ meVnm are obtained with $\mu=2.5$ meV while for $\alpha = 50$ meVnm with $\mu=2.3$ meV -- such they both correspond to the $\mu$ value for which
the topological transition occurs at the lowest $B$ -- see the dashed line in a).}
\label{decay_lengths}
\end{figure}

In Fig. \ref{decay_lengths} a) we plot map of the decay length $\xi$ in the topological phase $E_z^2 > \Delta^2 + \tilde{\mu}^2$ determined from the imaginary part of the wave-vector of the zero-energy solutions of the renormalized Hamiltonian Eq. (\ref{1DHamiltonian}) for a translational invariant system. Due to rescaling
of the chemical potential the contour of the topological regime in the diagram clearly deviates from the hyperbolic shape [see the inset to Fig. \ref{decay_lengths} a) for the phase diagram without the orbital effects] as recently measured for parallel orientation of the field \cite{chen_experimental_2017}.  With the white curve we depict the topological transition without the SO-paramagnetic effects, which shows narrower range of $B$ values for which the system is in the topological regime. 
The minimal decay length in the topological regime on the maps of Fig. \ref{decay_lengths} a) is significantly decreased from 41 nm for the case of neglected orbital effects, to 17 nm for the orbital effects included.

When the energy scale set by the SO coupling times the gap $\Delta$ is smaller than both the Zeeman energy $E_z$ and the chemical potential $\mu$ squared one can quantify \cite{das_sarma_splitting_2012} the Majorana decay length posterior the topological transition as,
\begin{equation}
\xi \simeq \frac{1}{\tilde{\alpha}\Delta}\sqrt{\left(\frac{\hbar^2}{\tilde{m}^*}\tilde{\mu}+\tilde{\alpha}^2\right)^2+\left(\frac{\hbar^2}{\tilde{m}^*}\right)^2\left(E_z^2-\Delta^2-\tilde{\mu}^2\right)}.
\label{xi_theory}
\end{equation}
Figure \ref{decay_lengths} b) with dashed curves shows the decay length for two strengths of the Rashba coupling  without the orbital effects of the magnetic field obtained through the above formula. The decay lengths with the orbital effect entering solely through the kinetic energy operator are plotted with the dot-dashed curves, while $\xi$ for the orbital effects included both in kinetic energy and SO coupling terms are depicted with solid curves. The behavior of the decay length obtained from Eq. (\ref{xi_theory}) reproduces the lengths inferred qualitatively from the band structure. The orbital effects through the kinetic energy operator decrease $\tilde{\mu}$ and $1/\tilde{m}^*$ and
by that reduce $\xi$, leading to the strong suppression of the decay length, up to a factor of three for the case of the strong SO coupling $\alpha =
50$ meVnm [confront the red dot-dashed curve with the dashed one]. The effects of SO-paramagnetic contribution are more convoluted. The increase of the magnetic field {\it increases}
the effective chemical potential $\tilde{\mu}$ through $E^{SO}_p$ term in Eq. (\ref{factormu}) counteracting the diamagnetic term contribution. On the
same time, the SO coupling constant $\tilde{\alpha}$ is decreased resulting in overall increase of $\xi \sim 1/\tilde{\alpha}$. 
This leads to a detrimental effect on the decay length reduction through the SO-paramagnetic effect which is manifested by the solid curves that
approach the dashed ones obtained with the sole Zeeman splitting. Note however that for the both considered Rashba strengths, $\xi$ still remains lower with
the orbital magnetic effects included as compared to the case of the Zeeman splitting only. 

The crosses in Fig. \ref{decay_lengths} b) corresponds to numerically obtained decay lengths from the renormalized Hamiltonian Eq.
(\ref{1DHamiltonian}). We observe excellent agreement of the two approaches. The only difference is the divergence of the decay length just after the
phase transition, where the decay length is dictated by the gapped Dirac cone at $k=0$ and which is not captured by the formula Eq. (\ref{xi_theory}).

\subsection{Beyond low-energy approximation.}
The above consideration focused on the band mixing limited only to the two lowest-energy transverse subbands. For stronger magnetic field or SO
coupling this approach must unavoidably break. Furthermore the validity of decay lengths obtained through Eq. (\ref{xi_theory}) is limited to small
$\Delta$ and $\alpha$ \cite{das_sarma_splitting_2012}. Now we turn our attention to exact solution of the Hamiltonian (\ref{2DHamiltonian}) to test the
validity of the developed theory and to extend our study beyond the above mentioned limits. For this purpose, we diagonalize numerically the
Hamiltonian (\ref{2DHamiltonian}) on a square mesh with $\Delta x = \Delta y = 4$ nm. The orbital effects of the magnetic field are incorporated using
Peierls substitution of the hopping elements $t_{nm} \rightarrow t_{nm}\exp\left[-ie\int\mathbf{A}d\mathbf{l}/\hbar\right]$.
In Fig. \ref{dispersions} c) with the black crosses we depict the dispersion relation obtained from the numerical calculations. As we see the agreement between the analytical and the numerical results is exceptionally good. 

\begin{figure}[ht!]
\center
\includegraphics[width = 7.5cm]{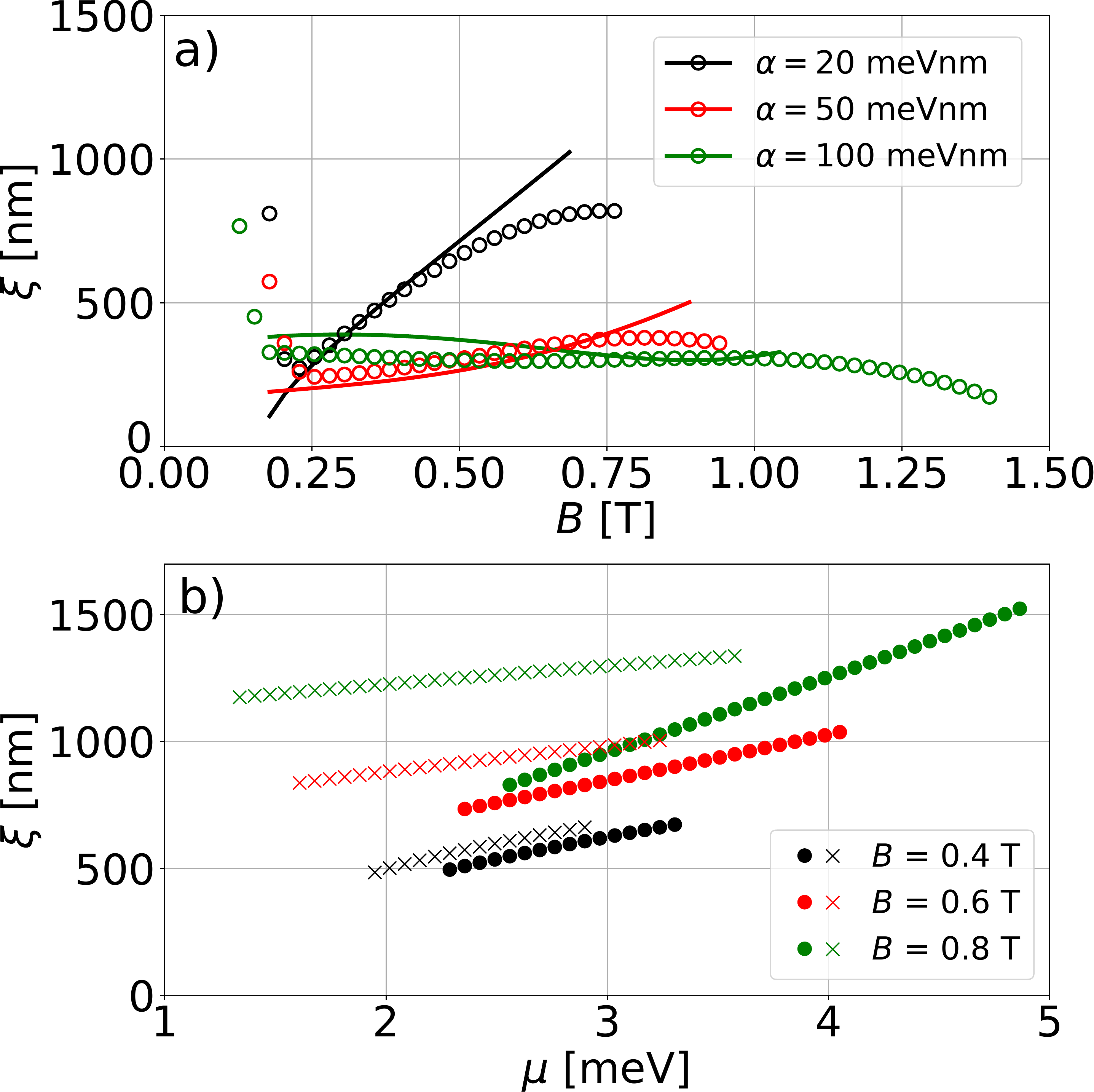}
\caption{a) Majorana bound state decay length $\xi$ as a function of the magnetic field for three values of the Rashba coupling strength. Solid curves
correspond to the analytical model of Eq. (\ref{xi_theory}) while the open circles are results of the exact numerical diagonalization of Hamiltonian
(\ref{2DHamiltonian}). Results for $\alpha=20,50$ meVnm correspond to those on Fig. \ref{decay_lengths} b) while the data for $\alpha=100$ meVnm is
obtained with $\mu=1.57$ meV. b) The decay length $\xi$ versus the chemical potential $\mu$ in the absence (the crosses) and presence (the circles) of the orbital effects for $\alpha = 20$ meVnm obtained through the Hamiltonian (\ref{2DHamiltonian}).
}
\label{decay_lengths_comparison}
\end{figure}

Finally, we complement the study with numerical assessment of the decay lengths. The spatial extent of MBSs is obtained numerically as the largest
decay length $\xi = \max \mathit{Re}[\kappa]^{-1}$ of the evanescent
waves $\Psi \sim e^{-\kappa x}$ at zero energy, with $\kappa$ being the eigenvalue of the translational operator \cite{sticlet_robustness_2017}. We
perform eigendecomposition of the translation operator for a infinite wire described by the Hamiltonian (\ref{2DHamiltonian}). The topological
transition is determined from the change of the topological invariant calculated as the determinant of the reflection
matrix $\mathrm{sgn}(\mathrm{det}[R])$ of a finite, translation invariant slice of the wire contacted with infinite
normal and superconducting electrodes \cite{fulga_scattering_2011}. All calculations are performed in Kwant package \cite{groth_kwant:_2014}.

Figure \ref{decay_lengths_comparison} a) with open circles presents numerical results for the decay lengths. We observe great agreement with the
developed theory  that persist up to the topological phase closing, where the increasing magnetic field and wave-vector breaks the applicability of
both the two-band model and validity of the formula given by Eq. (\ref{xi_theory}). Interestingly, in this regime, in the exact calculation we observe further beneficial reduction of the decay length which leads to decreasing amplitude of the energy oscillation of the overlapping MBSs [see Appendix C].

Inspecting the decay length as a function of the chemical potential we observe that the tunability of the $\xi$ is much more pronounced with the orbital effects included -- see the circles in Fig. \ref{decay_lengths_comparison} b). This is understandable since now the magnetic field modifies $\xi$ not only through $E_z$ but also by all the renormalized parameters $1/\tilde{m}^*$,  $\tilde{\mu}$,  $\tilde{\alpha}$ in Eq. (\ref{xi_theory}).

\section{Conclusions}
Summarizing, we studied impact of the orbital effects of the perpendicular magnetic field on the decay length of the Majorana bound states in the
proximitized semiconductor wire. We provided quasi-one-dimensional analytical model that allows to quantify the energies and decay lengths of Majorana
modes in the low-density limit, which we validated by comparison with exact numerical calculations. We found that the reduction of the decay length
via diamagnetic rescaling of the chemical potential is assisted by the change of the effective mass due to subband mixing by the paramagnetic term in
the kinetic energy operator. On the other hand, the vector potential entering through the SO coupling Hamiltonian has rudimentary effect on the decay
length reduction by the enhancement of the chemical potential and the decrease of the Rashba coupling constant. We found however, that in total, the
spatial extent of the Majorana modes is still less than without the orbital magnetic effects in favor of the topological robustness of 
Majorana states in finite size quantum gate devices on composite nanowires \cite{plissard_formation_2013, fadaly_observation_2017, gazibegovic_epitaxy_2017}.

\begin{acknowledgements}
This work was supported by National Science Centre, Poland (NCN) according to decision DEC-2016/23/D/ST3/00394. PW acknowledges support from the Faculty of Physics
and Applied Computer Science AGH UST statutory tasks within subsidy of Ministry of Science and Higher Education. The calculations were performed on PL-Grid Infrastructure.
\end{acknowledgements}

\appendix
\section{Derivation of the analytic formula for E(k) with the orbital effects}
In this subsection we demonstrate in details the folding-down procedure which leads to a single-band model with effective mass, chemical potential
and spin-orbit coupling renormalized due to the orbital effects. In the presence of a magnetic field $B$ the orbital effects couple spatial
degrees of freedom in the direction perpendicular to $B$. 
Bearing in mind that Majorana bound states are formed in the ground state of transverse quantization, in the low-energy approximation the system can be described in the framework of the two-band Hamiltonian. 
For this purpose we write the Hamiltonian (1) from the main paper, in a basis of the two lowest eigenstates of infinite quantum well in
the $y$-direction, $\Psi_n(y) = \sqrt{2/W}\sin(n\pi(y+W/2)/W)$. We obtain 
\begin{equation}
H=\left(
\begin{matrix}
H_{11} & H_{12} \\
H_{21} & H_{22}
\end{matrix}\right),
\label{A_two_band_model}
\end{equation}
with the diagonal elements given by
\begin{widetext}
\begin{equation}
 H_{11(22)}= \left (
\begin{matrix}
\frac{\hbar ^2 k_x ^2}{2m}-\mu_{1(2)}+\frac{1}{2}g\mu _B B	 & \Delta & i\alpha k_x & 0 \\
\Delta & -\frac{\hbar ^2 k_x ^2}{2m}+\mu _{1(2)}+\frac{1}{2}g\mu _B B & 0 & -i\alpha k_x \\
-i\alpha k_x & 0 &  \frac{\hbar ^2 k_x ^2}{2m}-\mu _{1(2)}-\frac{1}{2}g\mu _B B  & \Delta  \\
0 & i \alpha k_x & \Delta & -\frac{\hbar ^2 k_x ^2}{2m}+\mu_{1(2)} -\frac{1}{2}g\mu _B B 
\end{matrix}\right),
 \end{equation}
\end{widetext}
where $\alpha$ is the Rahsba spin-orbit coupling constant, the chemical potential $\mu _n=\mu-E_n-H_n^{dia}$ with  $n=1,2$, where  $E_n=
n^2\pi^2\hbar^2/2mW^2$ is the energy excitation in the $y$-direction. $H^{dia}_n=\langle \Psi_n |y^2| \Psi_n \rangle e^2B^2/2m$ is the diamagnetic
term in the $n$'th subband with $\langle \Psi_1 |y^2| \Psi_1 \rangle=(\pi^2-6)W^2/12\pi^2$ and $\langle \Psi_2 |y^2| \Psi_2 \rangle=
(2\pi^2-3)W^2/24\pi^2$, respectively, where $| \Psi_n \rangle = ( \psi _{n}^{e \uparrow},  \psi _{n}^{h \downarrow}, \psi _{n}^{e \downarrow}, - \psi
_{n}^{h \uparrow})$  \\
The off-diagonal elements of (\ref{A_two_band_model}) have the form
\begin{widetext}
\begin{equation}
 H_{12}= H_{21}=\left (
\begin{matrix}
\varepsilon_{p}k_x & 0 & -iE^{SO}_{p}+E^{SO}_\bot & 0 \\
0 & \varepsilon_{p}k_x & 0 & -i E^{SO}_{p}-E^{SO}_\bot \\
i E^{SO}_{p}+E^{SO}_\bot & 0 & \varepsilon_{p}k_x & 0 \\
0 & i E^{SO}_{p}-E^{SO}_\bot & 0 & \varepsilon_{p}k_x
\end{matrix}\right),
\end{equation}
where $\varepsilon_{p}=-\langle \Psi_1 | y| \Psi_2 \rangle e B \hbar /m =16W e B \hbar / 9 m \pi^2$, $E_{p}^{SO}=\langle\Psi_1|y|\Psi_2 \rangle\alpha
e B/\hbar = -16W\alpha e B /9\pi^2\hbar$ and $E_{\bot}^{SO} = i\alpha \langle\Psi_1 |\partial/\partial_x |\Psi_2\rangle = 8 i\alpha / 3W$.
\end{widetext}

\noindent Using the folding-down transformation, 
\begin{equation}
\label{A_eq:Hc}
 \mathcal{H}(E)=H_{11}-H_{12}(H_{22}-E)^{-1}H_{21}
\end{equation}
the $8\times 8$ Hamiltonian (\ref{A_two_band_model}) can be reduced into the $4\times 4$ effective Hamiltonian.
Based on the fact that $E_2$ is the largest energy in the system $E_2 \ll (\alpha^2 m^*/2\hbar^2, \Delta, 1/2 g \mu _B B)$, we can neglect the
spin-orbit coupling, superconducting pairing and the Zeeman splitting in the first excited state. Then, $(H_{22}-E)^{-1}$ takes the diagonal form 
\begin{equation}
 (H_{22}-E)^{-1}= \left [ \left ( \frac{\hbar ^2 k_x ^2}{2m}-\mu+E_2+H^{dia}_2 \right ) \sigma_0\tau_z -E \sigma_0\tau_0 \right ] ^{-1}.
\end{equation}
This expression can be further simplified by expanding each of the diagonal element into the Taylor series around $E_2$ and limit only to the first
term,
\begin{equation}
\begin{split}
&\frac{1}{ \frac{\hbar ^2 k_x ^2}{2m}-\mu+E_2+H^{dia}_2 -E} = \\
&\frac{1}{E_2}-\frac{1}{E_2^2} \left ( \frac{\hbar ^2 k_x ^2}{2m}-\mu+H^{dia}_2 -E 
\right ) + \cdots .
\end{split}
\end{equation}
As we see in the main paper, this approximation works well and the calculated dispersions $E(k)$ remain in a good agreement with the results obtained
from the exact diagonalization of Eq. (\ref{A_two_band_model}).
Based on the calculated term 
\begin{widetext}
\begin{equation}
 H_{12}(H_{22}-E)^{-1}H_{21}=\left (
\begin{matrix}
\frac{ \varepsilon_{p}^2 k_x^2}{E_2}+\frac{(E_{p}^{SO}-E_{\bot}^{SO})^2}{E_2} & 0 & -2i \frac{E^{SO}_{p}\varepsilon_{p}}{E_2} k_x & 0 \\
0 & -\frac{\varepsilon_{p}^2 k_x^2}{E_2}-\frac{(E_{p}^{SO}-E_{\bot}^{SO})^2}{E_2} & 0 & 2i \frac{E^{SO}_{p} \varepsilon_{p}}{E_2}k_x \\
2i \frac{E^{SO}_{p}\varepsilon_{p}}{E_2} k_x & 0 & \frac{\varepsilon _{p}^2 k_x^2}{E_2}+\frac{(E_{p}^{SO} - E_{\bot}^{SO})^2}{E_2} & 0 \\
0 & -2i \frac{E^{SO}_{p} \varepsilon_{p}}{E_2} k_x & 0 & -\frac{\varepsilon_{p}^2 k_x^2}{E_2}-\frac{(E_{p}^{SO}-E_{\bot}^{SO})^2}{E_2}
\end{matrix}\right),
\end{equation}
\end{widetext}
the folding-down procedure leads to
\begin{equation}
\mathcal{H} = \left( \hbar^2 k_x^2/2\tilde{m}^*- \tilde{\mu}\right) \sigma_0\tau_z+\Delta\sigma_0\tau_x 
-\tilde{\alpha} k_x\sigma_y\tau_z+E_Z\sigma_z\tau_0,
\label{A_ham_eff}
\end{equation}
where $\tilde{m}^*$, $\tilde{\mu}$, $\tilde{\Delta}_{SO}$ are the effective mass, chemical potential and SO coupling energy normalized due to the presence of
the orbital effects 
\begin{eqnarray}
\label{A_factors}
 \frac{1}{\tilde{m}^*}&=&\frac{1}{m^*}-\frac{2\varepsilon_{p}^2}{\hbar ^2 E_2},\\
 \tilde{\mu}&=&\mu-E_1-E_{1}^{dia}+\frac{(E^{SO}_p-E_{\bot}^{SO})^2}{E _2}, \\
 \tilde{\alpha}&=&\alpha  +2\frac{E^{SO}_p\varepsilon_{p}}{E _2}.
\end{eqnarray}

By diagonalization of Eq. (\ref{A_ham_eff}) we find the energies
\begin{equation}
\begin{split}
\label{A_analytical_energy}
&E(k_x) =\\ 
&\pm \sqrt{\tilde{\xi} _1^2+E_Z^2+\tilde{\Delta} _{SO}^2+\Delta ^2 
 \pm 2 \sqrt{E_Z^2\tilde{\xi}_1^2+\tilde{\Delta} _{SO}^2\tilde{\xi}_1^2+\Delta^2E_Z^2} },
\end{split}
\end{equation}
where $\tilde{\xi}_1=\hbar^2 k_x^2/2\tilde{m}^*-\tilde{\mu} $ and $\tilde{\Delta}_{SO}=\tilde{\alpha}k_x$.

\section{Impact of the nanowire width}
\begin{figure}[ht!]
\center
\includegraphics[width = 7 cm]{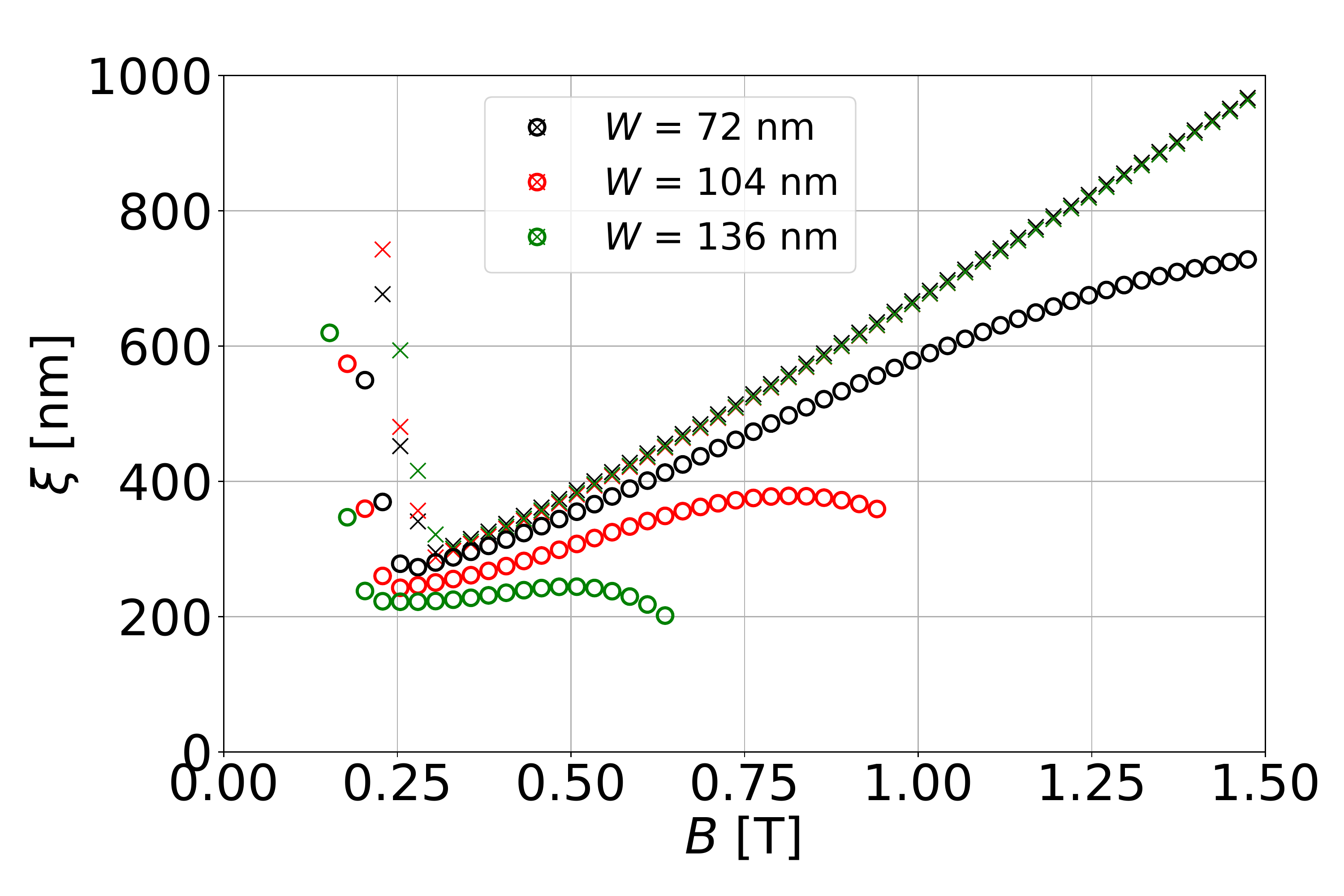}
\caption{Decay length $\xi$ versus the magnetic field $B$ for three values of the nanowire width for the orbital effects included (neglected) with circles (crosses). Results obtained in numerical calculation for  $\alpha = 50$ meVnm and $W=72$
nm -- $\mu=5$ meV, $W=104$ nm -- $\mu = 2.3$ meV, $W=136$ nm -- $\mu = 1.3$ meV.}
\label{decay_length_vs_B_W}
\end{figure}

We inspect the impact of the wire width on the MBSs decay length. Without the orbital effects the nanowire width does not affect $\xi$ as
can be observed in Fig. \ref{decay_length_vs_B_W}, provided that we tune the chemical potential such the phase transition occurs for the minimal $B$
in the phase diagram. Inclusion of the orbital effects significantly decreases the decay length at the cost of reduction of the topological phase size
in $B$.

\section{Energy spectra}
\begin{figure}[h!]
\center
\includegraphics[width = 6.5 cm]{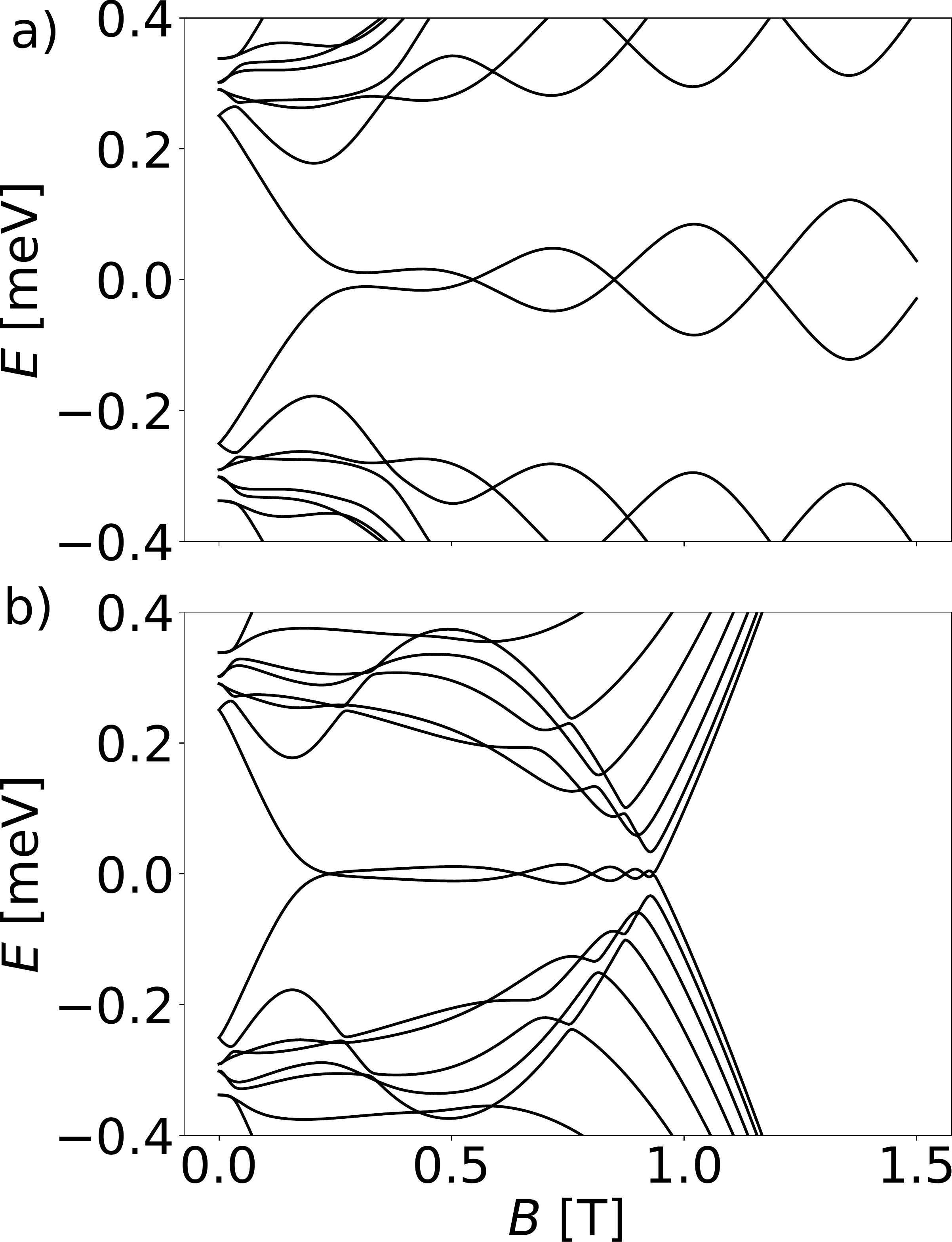}
\caption{Numerically obtained energy spectra without a) and with b) orbital effects included. Results for $\mu = 2.3$ meV, $W=104$ nm, $\alpha = 50$ meVnm calculated for a
finite system with the length $L=1000$ nm.}
\label{spectra}
\end{figure}
Without the orbital magnetic effects the energy of overlapping MBSs in a finite system deviates from zero, oscillating with a growing amplitude
when $B$ increases accordingly to Eq. (\ref{xi_theory}) of the main text. Inclusion of the orbital effects decreases $\xi$ near the phase transition at high magnetic fields and through that limits the energy oscillations of MBSs, as can be observed in Fig. \ref{spectra}. 

\bibliography{orbital}
\end{document}